\documentclass[PRC,aps,noshowpacs,twocolumn,superscriptaddress,floatfix]{revtex4-1}
\usepackage{graphicx}
\usepackage{epsfig}
\usepackage{amsfonts}
\usepackage{amsmath,amssymb}
\usepackage{bm}

\usepackage[usenames]{color}
\usepackage{ulem}

\def\bea{\begin{eqnarray}}
\def\eea{\end{eqnarray}}
\def\be{\begin{equation}}
\def\ee{\end{equation}}

\def\lsim{~\,\makebox(1,1){$\stackrel{<}{\widetilde{}}$}\,~}

\begin{document}

\preprint{APT/123-QED
\hspace{0.2cm}
ZTF-EP-14-04}

\title{Determination of symmetry energy from experimental and observational constraints; 
prediction on CREX}

\author{Shingo Tagami}
\affiliation{Department of Physics, Kyushu University, Fukuoka 812-8581, Japan}

\author{Nobutoshi Yasutake}
\affiliation{Department of Physics, Chiba Institute of Technology,
             Chiba 275-0023, Japan}
\affiliation{Advanced Science Research Center, Japan Atomic Energy Agency, Ibaraki 319-1195, Japan}

\author{Mitsunori Fukuda}
\affiliation{Department of Physics, Osaka University, Osaka 560-0043, Japan}

\author{Jun~Matsui}
\affiliation{Department of Physics, Kyushu University, Fukuoka 812-8581, Japan}

\author{Masanobu Yahiro}
\email[]{orion093g@gmail.com}
\affiliation{Department of Physics, Kyushu University, Fukuoka 812-8581, Japan}

\date{\today}

\begin{abstract}
Taking $r_{\rm skin}^{208}({\rm PREX})=0.33^{+0.16}_{-0.18}~{\rm fm}$ as an experimental constraint and $M_{\rm max}^{\rm NS} \ge 2{\rm M}_{\rm sun}$ as an observational (astrophysical) constraint,  
we determine an indisputable range for $J$, $L$, $K_{\rm sym}$ defined in Eq.~\eqref{eq-S-sym}. 
For this purpose, we take a statistical approach. We first accumulate the 206 EoS data from theoretical works and 
take correlation between $r_{\rm skin}^{208}$ and $L$ for the 206 EoSs, where 7 Gogny EoSs are 
obtained by our calculations.    
Since the correlation coefficient is $R = 0.99$, we can regard $L$ as a function of $r_{\rm skin}^{208}$, so that
we succeed in deducing an empirical constraint  $L=31-161$~MeV 
from  $r_{\rm skin}^{208}({\rm PREX})=0.15-0.49$~fm.    
For the 47 EoSs satisfying the observational constraint, 46 EoSs satisfy the empirical constraint. 
The 46 EoSs yield $J=29-44$~MeV, $L= 37-135$~MeV, $K_{\rm sym}=(-137)-(160)$~MeV.  
The is a primary result.  
When we take correlation between $r_{\rm skin}^{48}$ and $r_{\rm skin}^{208}$ for the 206 EoSs, 
$R$ is 0.99. 
The  $r_{\rm skin}^{48}$--$r_{\rm skin}^{208}$ relation allows us to transform 
$r_{\rm skin}^{208}({\rm PREX})$ into the corresponding data on $r_{\rm skin}^{48}$
In order to estimate a value of $r_{\rm skin}^{48}$ for ongoing CREX, we take the weighted mean and its error of 
two present data on $r_{\rm skin}^{48}$ and transformed PREX value on $r_{\rm skin}^{48}$. The weighted mean is 
$r_{\rm skin}^{48}=0.17$~fm. This is a prediction for the central value of CREX.   
\end{abstract}

\maketitle

\section{Introduction and Conclusion}
\label{sec-Introduction}
{\it Background:}
For the symmetry energy $S_{\rm sym}(\rho)$ as 
a function of nuclear density $\rho$, many predictions have been made so far.  
The $S_{\rm sym}(\rho)$ is expanded around the nuclear-matter saturation density $\rho_0$:    
\bea
S_{\rm sym}(\rho)=J + \frac{L (\rho-\rho_0)}{3\rho_0}+\frac{K_{\rm sym} (\rho-\rho_0)^2}{18\rho_0^2}+ 
\cdots 
\label{eq-S-sym}
\eea  
for the symmetry energy $J \equiv S_{\rm sym}(\rho)$, the slope $L$ and the curvature $K_{\rm sym}$ at $\rho=\rho_0$. 
The $S_{\rm sym}(\rho)$ were determined from some experimental and observational constraints 
and their combinations. It is important which constraint is selected, particularly for  neutron stars (NSs).

The  $S_{\rm sym}(\rho)$ itself cannot be measured by experiments. 
In place of  $S_{\rm sym}(\rho)$, the neutron-skin thickness $r_{\rm skin}$ is measured to determine 
$L$, since the strong correlation between $r_{\rm skin}^{208}$ and $L$ is well known~\cite{RocaMaza:2011pm,Brown:2013mga}. 
The direct measurement on $r_{\rm skin}^{208}$ is the Lead Radius EXperiment (PREX) 
composed of  parity-violating and elastic electron scattering; 
the neutron radius $r_{\rm n}$ is determined from the former experiment, whereas
the proton radius $r_{\rm p}$ is from the latter. 
The weak and electromagnetic measurement provides $r_{\rm n}^{208}=5.78^{+0.16}_{-0.18}$~fm~\cite{PREX05,Abrahamyan:2012gp} that yields 
\bea
r_{\rm skin}^{208}{\rm (PREX)}=0.33^{+0.16}_{-0.18}~{\rm fm} 
\label{Eq:PREX}
\eea
under $r_{\rm p}^{208}=5.45$~fm.
The $r_{\rm skin}^{208}{\rm (PREX)}$ has a large error. For this reason, 
the PREX-II and the $^{48}$Ca Radius EXperiment (CREX) are ongoing at Jefferson Lab~\cite{PREX05}. 
As for $^{48}$Ca, Hagen  {\it et al.} obtained a value of {proton radius} $r_p$ 
with the coupled-cluster calculation with the chiral interaction~\cite{Hagen:2015yea}. Using the $J$--$r_p$ and $L$--$r_p$ correlations, they have showed $25.2 \lsim J \lsim 30.4$ MeV and $37.8 \lsim L \lsim 47.7$ MeV, where 
the theoretical errors of $J$ and $L$ mainly come from the systematic uncertainties of the employed Hamiltonians.

The $S_{\rm sym}(\rho)$ influences strongly the nature within NSs. For a pulsar in a binary system, detection of the general relativistic Shapiro delay allows us to determine the mass $M$ of NS. 
In fact, the NS with $M = 1.97 \pm 0.04~{\rm M}_{\rm sun}$ is observed~\cite{Demorest:2010bx}. 
In $\rho =2-3 ~\rho_{0}$, 
the strangeness changing weak decay replaces nucleons by hyperons. 
The decay makes the maximum mass $M_{\rm max}$ decrease, if the decay occurs really. 
In this paper,  we assume the decay does not take place, because nucleon-hyperon and hyperon-hyperon interactions are still unknown. 
As an essential constraint on the EoS from astrophysics, one can consider  
\bea
M_{\rm max} \ge 2~{\rm M}_{\rm sun},   
\label{Eq:2solar mass constraint-0}
\eea
 where $M_{\rm max}$ is  the maximum mass of NS and  ${\rm M}_{\rm sun}$ is the solar mass.

{\it Aims and conclusion:}
The first aim is to  determine an indisputable range for $J$, $L$ and $K_{\rm sym}$ 
from  $r_{\rm skin}^{208}({\rm PREX})=0.33^{+0.16}_{-0.18}~{\rm fm}$ as an experimental constraint and $M_{\rm max}^{\rm NS} \ge 2{\rm M}_{\rm sun}$ as an observational (astrophysical) constraint. 
Our result is $J=29-44$~MeV, $L= 37-135$~MeV, $K_{\rm sym}=(-137)-(160)$~MeV. 

For the first aim, we take a statistical approach.
We first take a  correlation between $r_{\rm skin}^{208}$ and  $L$. 
The number of EoSs should be large for obtaining reliable correlation statistically. 
We then accumulate 204 EoSs from theoretical works and construct 2 Gogny-type EoSs, so that 
we get 206 EoSs.  
An an advantage of this statistical analysis is to eliminate systematic error for the correlation between 
$r_{\rm skin}^{208}$ and  $L$. 
The procedure of the statistical analysis is as follows. 
\begin{enumerate}

   \item 
 We take Eq.~\eqref{Eq:PREX} as  an essential experimental constraint and 
 Eq.~\eqref{Eq:2solar mass constraint-0} as  an essential observational constraint.

  \item 
In order to take reliable correlation between  $r_{\rm skin}^{208}$ and  $L$, 
we accumulate 204 EoS data from theoretical works~\cite{Akmal:1998cf,RocaMaza:2011pm,Ishizuka:2014jsa,Gonzalez-Boquera:2017rzy,D1P-99,Gonzalez-Boquera:2017uep,Oertel:2016bki,Piekarewicz:2007dx,Lim:2013tqa,Sellahewa:2014nia,Inakura:2015cla,Fattoyev:2013yaa,Steiner:2004fi,Centelles:2010qh,Dutra:2012mb,Brown:2013pwa,Brown:2000pd,Reinhard:2016sce,Tsang:2019ymt,Ducoin:2010as,Fortin:2016hny,Chen:2010qx,Zhao:2016ujh,Zhang:2017hvh,Wang:2014rva,Lourenco:2020qft} and construct 2 Gogny-type EoSs 
 in which $r_{\rm skin}^{208}$ and/or $L$ is presented. 
  The 206 EoSs are presented in Table I. 
 The correlation function between $r_{\rm skin}^{208}$ and $L$ is obtained self-consistently; 
 the starting correlation function is  determined from the EoSs 
 in which $r_{\rm skin}^{208}$ and $L$ are provided. 
 The resulting correlation function has $R=0.99$, while the  starting correlation function does $R=0.98$. 
 Since the resulting correlation is perfect, we refer to the resulting correlation as 
 ``the $r_{\rm skin}^{208}$--$L$ relation''. In other words, $L$ is a function of $r_{\rm skin}^{208}$; 
 see Fig.~\ref{fig:Skin-Pb208-L-1}. 
 Using the $L$--$r_{\rm skin}^{208}$ relation, we get an empirical constraint $L_{\rm empirical}=31-161$ 
 from $r_{\rm skin}^{208}{\rm (PREX)}$.

 \item 
For 7 Gogny-typ EoSs in Table I,  both $r_{\rm skin}^{208}$ and $L$ are not calculated. We then calculate $J$, $L$, $K_{\rm sym}$ with the energy density functional 
and $r_{\rm skin}^{48}$ and $r_{\rm skin}^{208}$ with the Hartree-Fock-Bogoliubov method 
with the angular momentum projection. Two of 7 EoSs are newly constructed.

  \item 
 The 184 EoSs satisfying $L_{\rm empirical}=31-161$ in Table I yield 
 $J=28-44$~MeV, $L= 31-161$~MeV, $K_{\rm sym}=(-266)-(235)$~MeV. 
 
  \item 
  Table III shows the 47 EoSs satisfying the observational constraint ~\eqref{Eq:2solar mass constraint-0}. 
  46 EoSs satisfying $L_{\rm empirical}=31-161$ in Table III give  
 $J=29-44$~MeV, $L= 37-135$~MeV, $K_{\rm sym}=(-137)-(160)$~MeV. 
 These are indisputable ranges of  $J$, $L$, $K_{\rm sym}$. 
  
\end{enumerate}

The second aim is  to  estimate a central value of CREX. 
We find that the correlation between $r_{\rm skin}^{48}$ and $r_{\rm skin}^{208}$ is perfect, 
because $R=0.99$. 
The  $r_{\rm skin}^{48}$--$r_{\rm skin}^{208}$ relation allows us to transform $r_{\rm skin}^{208}{\rm (PREX)}$ 
into the corresponding range on $r_{\rm skin}^{48}$. We then  take the weighted mean and its error of 
two data~\cite{Birkhan:2016qkr,Tanaka:2019pdo} measured for $r_{\rm skin}^{48}$ and the transformed  $r_{\rm skin}^{208}{\rm (PREX)}$. The weighted mean is a prediction on the central value of CREX; namely, 
\bea
r_{\rm skin}^{48}({\rm CREX})  =0.17~{\rm fm}.
\eea.

We show the statistical analysis  and its results in Sec.~\ref{Sec:The statistical analysis and its results} and 
our estimate for the central value of CREX in Sec.~\ref{Sec:Prediction for CREX}. 
Discussions are made in Sec.~\ref{Sec:Discussions}. 

\section{The statistical analysis and its results}
\label{Sec:The statistical analysis and its results}

\subsection{Sample data of 206 EoSs}
We first accumulate 204 EoSs from theoretical works~\cite{Akmal:1998cf,RocaMaza:2011pm,Ishizuka:2014jsa,Gonzalez-Boquera:2017rzy,D1P-99,Gonzalez-Boquera:2017uep,Oertel:2016bki,Piekarewicz:2007dx,Lim:2013tqa,Sellahewa:2014nia,Inakura:2015cla,Fattoyev:2013yaa,Steiner:2004fi,Centelles:2010qh,Dutra:2012mb,Brown:2013pwa,Brown:2000pd,Reinhard:2016sce,Tsang:2019ymt,Ducoin:2010as,Fortin:2016hny,Chen:2010qx,Zhao:2016ujh,Zhang:2017hvh,Wang:2014rva,Lourenco:2020qft} in which $r_{\rm skin}^{208}$ and/or $L$ is presented. 
In the 204 EoSs of Table I,  both $r_{\rm skin}^{208}$ and $L$ are not presented for 5 Gogny-type EoSs 
(D1S, D1N, D1M, D1M*, D1P).
We then construct 2 Gogny-type EoSs, D1MK and D1PK, in which strong three-body forces are into account. 
Eventually, we get the 206 EoSs, as shown in Table I. 
Now,  we make three comments for Table I.

\begin{enumerate}
  \item 
For D1S, D1N, D1M, D1M*, D1MK, D1P, D1PK, 
we calculate $r_{\rm skin}^{208}$ and $r_{\rm skin}^{48}$ 
with the Hartree-Fock-Bogoliubov method with the angular momentum projection. 
For the Gogny EoSs, the effective nucleon-nucleon interaction can be described as   
\bea
~~~~~~V(\vec{r})  &=&\sum_{i=1,2} t_0^i (1+x_0^i P_\sigma) \rho^{\alpha_i} \delta (\vec{r})
  \nonumber \\
 &+&\sum_{i=1,2} (W_i+B_i P_\sigma -H_i P_\tau -M_i P_\sigma P_\tau) e^{-\frac{r^2}{\mu_i^2}}
  \nonumber \\
 &+&i W_0 (\sigma_1 +\sigma_2) [\vec{k'} \times \delta (\vec{r}) \vec{k}], 
\eea 
where $\sigma$ and $\tau$ are the Pauli spin and isospin operators,
respectively, and the corresponding exchange operators $P_{\sigma}$ and $P_{\tau}$ are defined as usual.
See Table II for the parameter sets of D1MK and D1PK.  
For matter, the energy density functional is used.

  \item Dutra ${\it et~al.}$ selected 16 EoSs from 240 Skyrme EoSs by using 
a  series of criteria~\cite{Dutra:2012mb}. The 16 EoSs (
GSkI,
GSkII,
KDE0v1,
LNS,
MSL0,
NRAPR,
Ska25s20,
Ska35s20,
SKRA,
SkT1,
SkT2,
SkT3,
Skxs20,
SQMC650,
SQMC700,
SV-sym32)
 are in Table I.  
   
  \item 
Out of the 240 Skyrme EoSs, Tsang {\it et al.} selected 12 EoSs 
(KDE0v1,
NRAPR,
Ska25,
Ska35,
SKRA,
SkT1,
SkT2,
SkT3,
SQMC750,
SV-sym32,
SLy4,
SkM*
)
and fitted them to nuclear binding energies, charge radii 
and single-particle energies~\cite{Tsang:2019ymt}. The  fitted set is identified  
with the label ``-T'' from the original set.     
Brown and Schwenk fitted the original set so that the effective mass $m^*/m$ can be 0.9 
in neutron matter at $\rho = 0.10$~fm$^{-3}$~\cite{Brown:2013pwa}.  
This set  is identified with the label by ``-B''. 
The three sets are in Table I.  

\end{enumerate}

\subsection{$L$ as  a function of $r_{\rm skin}^{208}$}

As a correlation coefficient $R$ between two variables $X$ and $Y$, we take  
\bea
R=\frac{\sigma_{XY}}{\sigma_X \sigma_Y}
\eea
for the covariance $\sigma_{XY}$ between $X$ and $Y$ and the standard deviation $\sigma_{X}$ 
of $X$.

\begin{figure}[hbt]
\centering
\vspace{0cm}
\includegraphics[width=0.475\textwidth]{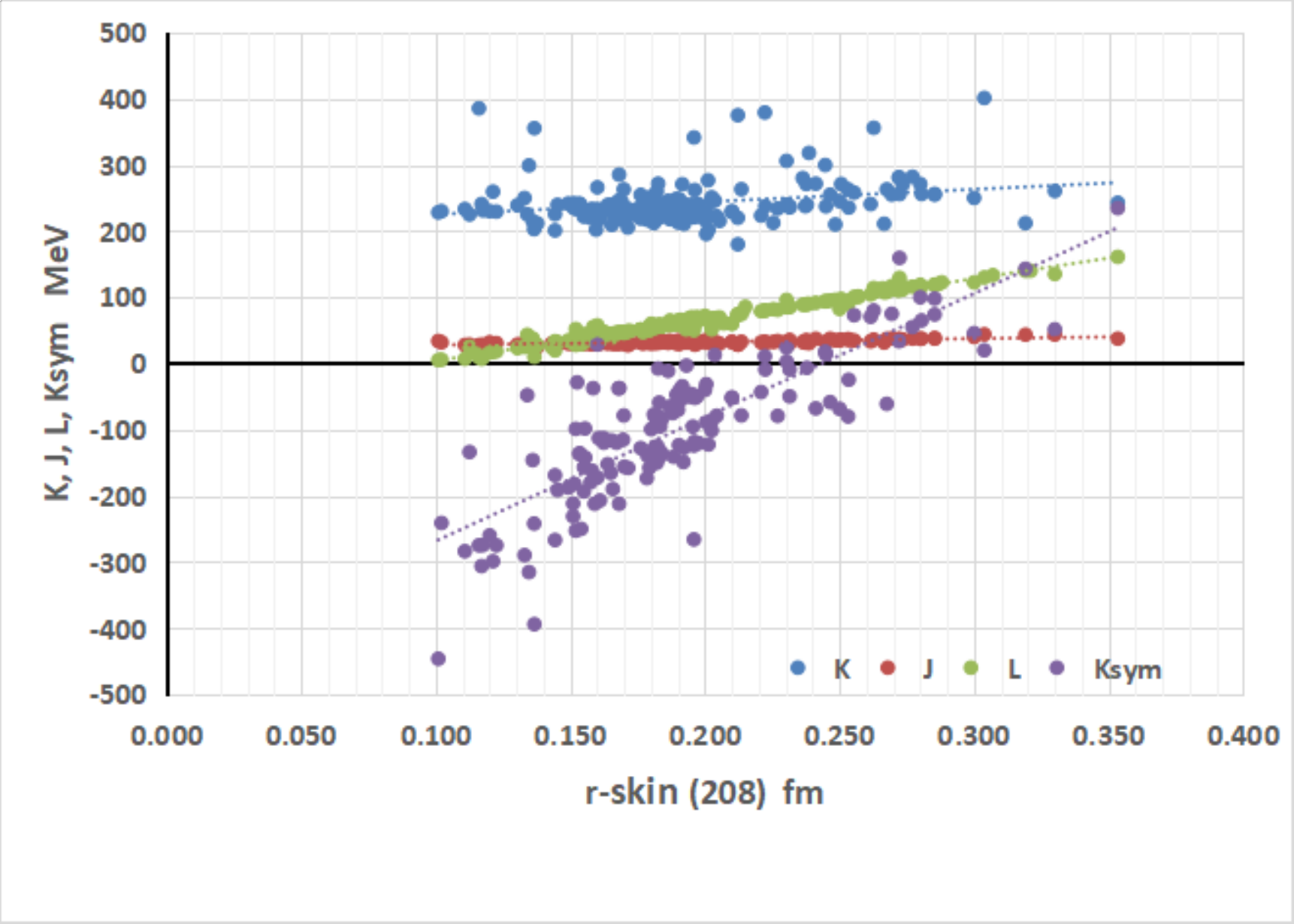}
\vspace{-10pt}
\caption{Properties ($J$, $L$, $K$, $K_{\rm sym}$) of 206 EoSs (dotts). 
Four dotted lines stand for correlations between $r_{\rm skin}^{208}$ and each of $J$, $L$, $K$, $K_{\rm sym}$. 
}
\label{fig:Skin-Pb208-L-1}
\end{figure}

Figure \ref{fig:Skin-Pb208-L-1} shows correlations between $r_{\rm skin}^{208}$ and each of 
$J$, $L$, $K$, $K_{\rm sym}$ for 206 EoSs  of Table~\ref{table-I}, where $K$ and $K_{\rm sym}$ are the incompressibility and the symmetry-energy incompressibility, respectively. 
Four dotted lines denote 
\bea
L=618.24~r_{\rm skin}^{208}-57.399 
\label{Eq:skin-L-starting}
\eea
for the correlation between $L$ and $r_{\rm skin}^{208}$ , 
\bea
J  =48.287~r_{\rm skin}^{208}+23.091
\label{Eq:skin-J}
\eea
for the correlation between $J$ and $r_{\rm skin}^{208}$, 
\bea
K_{\rm sym}  =1838.2~r_{\rm skin}^{208}-446.57
\label{Eq:skin-K-sim}
\eea
for the correlation between $K_{\rm sym}$ and $r_{\rm skin}^{208}$
\bea
K  =206.24~r_{\rm skin}^{208}+201.82
\label{Eq:skin-K}
\eea
for the correlation between $K_{\rm sym}$ and $r_{\rm skin}^{208}$, respectively. 
The correlation coefficient  is $R=0.98$ for $L$, $R=0.74$ for $J$, $R=0.84$ for $K_{\rm sym}$, 
$R=0.30$ for $K$. 
Among the four correlations, only the correlation \eqref{Eq:skin-L-starting}  
between $L$ and $r_{\rm skin}^{208}$ can be regarded as a function, since the $R$ is perfect.

In order to obtain both $L$ and $r_{\rm skin}^{208}$ for all of the 206 EoSs, 
we take a self-consistent way where the starting function is Eq.~\eqref{Eq:skin-L-starting}. 
The resulting function      
\bea
L{(r_{\rm skin}^{208})}=620.39~r_{\rm skin}^{208}-57.963 
\label{Eq:skin-L}
\eea
has $R=0.99$. We refer to Eq.~\eqref{Eq:skin-L} as ``the $r_{\rm skin}^{208}$--$L$ relation''.

\subsection{Empirical constraint on $L$}

Using the $r_{\rm skin}^{208}$--$L$ relation~\eqref{Eq:skin-L}, we can get an empirical constraint  
\bea
L_{\rm empirical}=31-161~{\rm MeV} 
\label{Eq:empirical range of L}
\eea  
from the experimental constraint $r_{\rm skin}^{208}{\rm (PREX)}$.

In Table I, 184 EoSs  satisfy the empirical constraint \eqref{Eq:empirical range of L} on $L$. 
For $J$, $L$, $K_{\rm sym}$, the 184EoSs yield  
\bea
J=28-44,~L=31-161,~K_{\rm sym}=(-266)-(235)~~~~~
\label{Eq:experimental result}
\eea
in units of MeV.

\subsection{Observational and empirical constraints}

In addition to the empirical constraint \eqref{Eq:empirical range of L} on $L$, we take 
the observational  constraint~\eqref{Eq:2solar mass constraint-0}. The procedure is as follows.

In Table I, 47 EoSs satisfy  
the observational constraint~\eqref{Eq:2solar mass constraint-0}; see Table III. 
Out of the 47 EoSs, the 46
 (
 BSk25,
BSk20,
BSk26,
D1M*,
Sly230a,
D2,
SLy4,
BSk24,
BSk21,
SFHo,
Sly2,
SKb,
DD-ME2,
Sly9,
D1PK,
DD-ME1,
HS(DD2),
NL3$\omega\rho$,
APR(E0019),
BSk23,
SkI6,
SkI4,
FSUgold2.1,
BSR2,
SGI,
D1AS,
BSk22,
SkMP,
LS220,
LS375,
SKa,
Rs,
TFa,
BSR6,
TMA(E0008),
SK272,
GM1,
SK255,
SV,
SkI3,
SkI2,
TM1,
NL3,
TFb,
SKI5,
TFc
)
satisfy the empirical constraint \eqref{Eq:empirical range of L} on $L$.  
We take $J$, $L$, $K_{\rm sym}$ from  the 46 EoS. 
The results are 
\bea
J=29-44,~L= 37-135,~K_{\rm sym}=(-137)-(160)~~~~~
\label{Eq:2solar mass result}
\eea
in units of MeV. This is our primary result for $J$, $L$, $K_{\rm sym}$, 
since we use Eq.~\eqref{Eq:PREX} 
as  an essential experimental constraint and 
Eq.~\eqref{Eq:2solar mass constraint-0} as  an essential observational constraint.  

\bigskip

\section{Prediction for CREX}
\label{Sec:Prediction for CREX}

\subsection{Experimental data on $r_{\rm skin}^{48}$}

At the present stage, there are two good measurements for $r_{\rm skin}^{48}$
The high-resolution measurement of $\alpha_{\rm D}$ was applied for $^{48}$Ca 
in RCNP~\cite{Birkhan:2016qkr}: the data  is
\bea
r_{\rm skin}^{48}{({\rm RCNP})}=0.14-0.20~{\rm fm}. 
\label{Eq:Experimental constraint-48}
\eea
Very lately, Tanaka {\it et al.}determined 
\bea
r_{\rm skin}^{48} ({\rm RIKEN}) =0.146 \pm 0.06~ {\rm fm}
\label{Eq:skin-Ca48-Tanaka} 
\eea
by measuring interaction cross sections for $^{48}$Ca  scattering on a C target 
in RIKEN~\cite{Tanaka:2019pdo}. 

\subsection{Relation between $r_{\rm skin}^{48}$ and $r_{\rm skin}^{208}$}

\begin{figure}[hbt]
\centering
\vspace{0cm}
\includegraphics[width=0.475\textwidth]{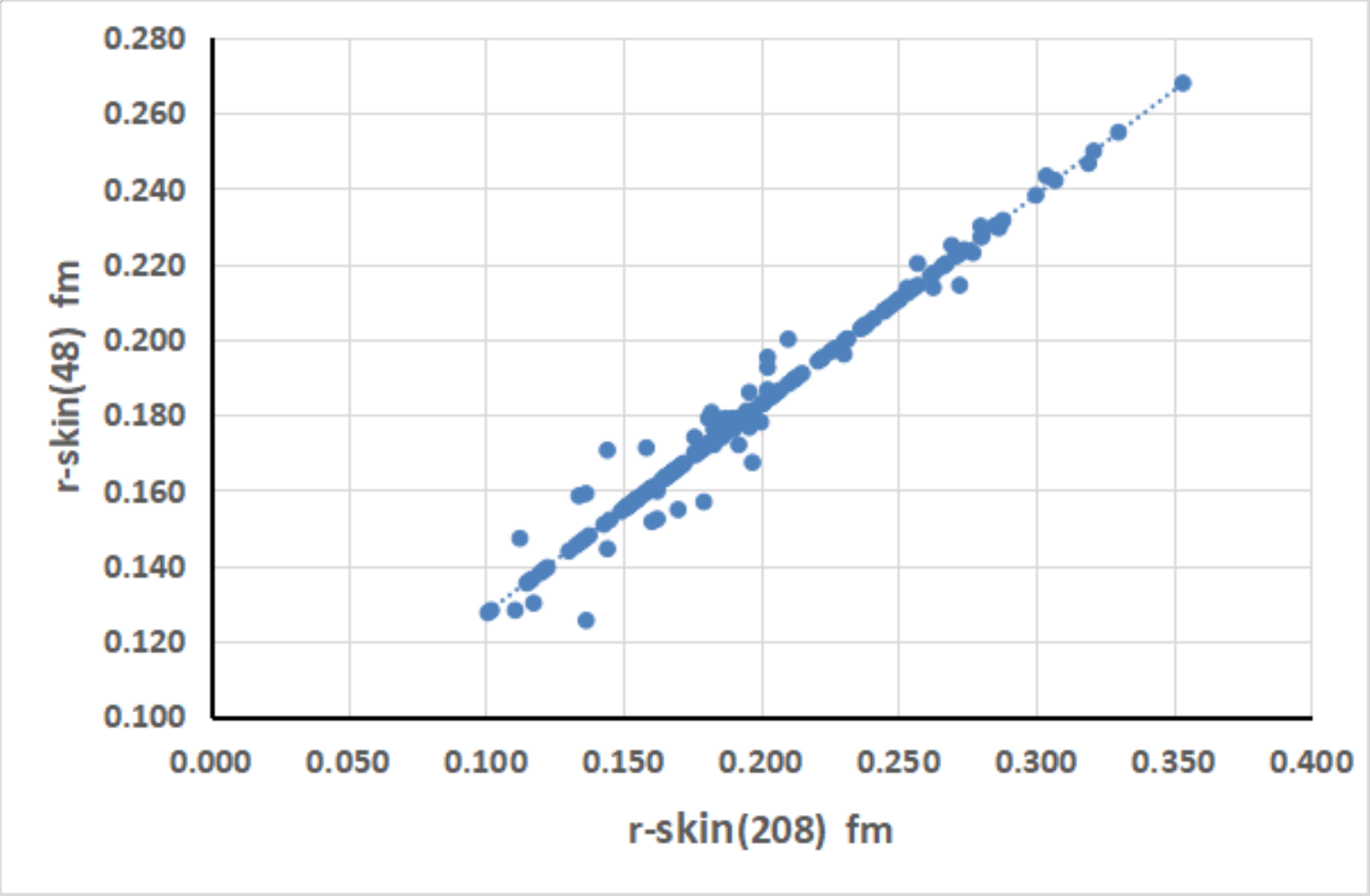}
\vspace{-10pt}
\caption{Relation between $r_{\rm skin}^{48}$ and $r_{\rm skin}^{208}$.  
The dotted line stands for Eq.~\eqref{Eq:skin-208-48-1}, while dots correspond to 206 EoSs. 
}
\label{fig:Skin-Pb208-Ca48-1}
\end{figure}

For the 206 EoSs in Table I, we take correlation between $r_{\rm skin}^{48}$ and $r_{\rm skin}^{208}$ 
for the EoSs in which both $r_{\rm skin}^{48}$ and $r_{\rm skin}^{208}$ are presented. 
The correlation coefficient is perfect because of $R=0.98$. In order to yield values of $r_{\rm skin}^{48}$ and $r_{\rm skin}^{208}$ for all of 206 EoSs, we take correlation 
between $r_{\rm skin}^{48}$ and $r_{\rm skin}^{208}$ self-consistently: 
the starting $r_{\rm skin}^{48}$--$r_{\rm skin}^{208}$ relation is  obtained from the EoSs 
in which $r_{\rm skin}^{48}$ and $r_{\rm skin}^{208}$ are presented. Since the resulting relation 
has $R=0.98$, we succeed in obtaining $r_{\rm skin}^{48}$ as a function of $r_{\rm skin}^{208}$. 
The function thus obtained is 
\bea
r_{\rm skin}^{48}(r_{\rm skin}^{208}) =0.5547~r_{\rm skin}^{208}+0.0718;
\label{Eq:skin-208-48-1}
\eea
see Fig.~\ref{fig:Skin-Pb208-Ca48-1}. 

\subsection{Data on $r_{\rm skin}^{48}$ transformed from $r_{\rm skin}^{208}{\rm (PREX)}$}

Transforming the data .$r_{\rm skin}^{208}{\rm (PREX)}$ into the corresponding data 
on $r_{\rm skin}^{48}$, we get 
\bea
r_{\rm skin}^{48}{\rm (tPREX)}  =0.25 \pm  0.09~ {\rm fm}. 
\label{Eq:skin-208to48} 
\eea
The transformed data \eqref{Eq:skin-208to48} is consistent with  the original ones~\eqref{Eq:Experimental constraint-48} and \eqref{Eq:skin-Ca48-Tanaka}.

Taking the weighted mean and its error of Eqs.~\eqref{Eq:Experimental constraint-48}, \eqref{Eq:skin-Ca48-Tanaka}, \eqref{Eq:skin-208to48}, we get 
\bea
r_{\rm skin}^{48}{\rm (RCNP ~\&~ RIKEN~\&~tPREX)}  =0.17 \pm 0.03~ {\rm fm}.~~~~~~ 
\label{Eq:skin-208to48} 
\eea 
We then regard 0.17~fm as an estimate for the central value of CREX. 
The derivation 0.03~fm is based on the ``maximum likelihood", in which we assume that the kernels of prior probabilities on the measurements are independent Gaussian types. The details are shown in Appendix.
Since the estimated error of CREX  is  0.02 fm~\cite{PREX05}, the CREX group 
will show their result with similar accuracy in future.

\section{Discussions}
\label{Sec:Discussions}

\subsection{Astrophysical constraints}
\label{Sec:Astrophysical constraints}

\begin{figure}[hbt]
\centering
\vspace{0cm}
\hspace{-5mm}
\includegraphics[width=0.5\textwidth]{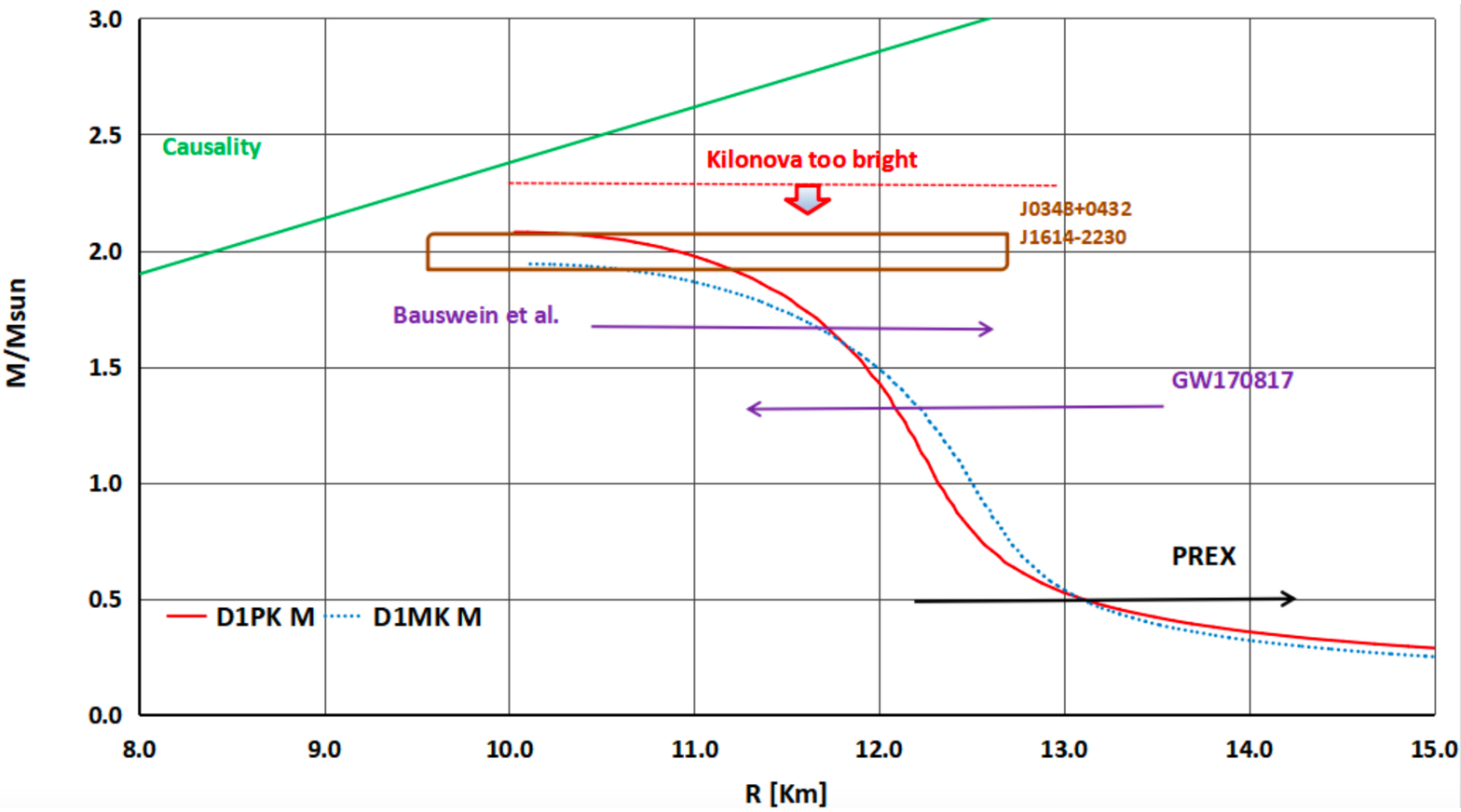}
\vspace{-35pt}
\caption{
$M$--$R$ relation for D1MK and D1PK. 
Model D1PK is shown by the solid line, and D1MK is by the dashed line. 
See the text for the observation constraints shown and 
the lower limit of $R$ determined from PREX. 
The details are shown Ref.~\cite{Horowitz:2019piw}.
}
\label{fig:M--R relation}
\end{figure}

As for the astrophysical constraints, 
Horowitz summarized the other observational constraints and the lower limit of NS radius $R$ determined 
from PREX~\cite{Horowitz:2019piw}; his mention is illustrated in Fig.~\ref{fig:M--R relation}, with the mass-radius relations for our EoSs, D1MK and D1PK. 
Bauswein {\it et al.} suggested that if 1.6 $M_{\rm sun}$ stars have radii less than the indicated
lower limit, the NS in GW170817 would collapse to soon to a black hole and not eject
material enough to power the observed Kilonova~\cite{Bauswein:2016}.
When the maximum mass of a NS is above the
red dotted line, Metzger {\it et al.} argued that the compact remnant in GW170817 lives
too long and provides too much energy to the kilonova~\cite{Metzger:2017}.
Finally the GW170817 limit on the radius of a 1.4 $M_{\rm sun}$ is from the limit on the gravitational deformability~\cite{LIGO, Fattoyev:2018}, whereas the limit on the EoS at low density from PREX is plotted as a minimum radius of a 0.5 $M_{\rm sun}$ NS. 
Such low mass NSs have low central densities comparable to nuclear density. 

In our calculations for the mass-radius relation, the beta-equilibrium is taken into account. Below the subnuclear density 
$n < 0.1$~fm$^{-3}$, we use the BPS EoS~\cite{Baym1971}.
Model D1PK satisfies the observable constraint~\eqref{Eq:2solar mass constraint-0}. 
Meanwhile, the maximum mass $M_{\rm max}=1.94~{\rm M_{sun}}$ of D1MK is slightly smaller than 
the observable constraint~\eqref{Eq:2solar mass constraint-0},  but is in the $M = 1.97 \pm 0.04~{\rm M}_{\rm sun}$ 
of Ref.~\cite{Demorest:2010bx}.
As an interesting result, Model D1PK satisfies all of the observation constraints mentioned above and 
the lower bound of $R$ determined from PREX. 
Model D1MK almost satisfies the observation constraints mentioned above and 
the lower bound of $R$ determined from PREX.

\begin{acknowledgments}
M. Y. expresses my  gratitude to  Dr. Y. R. Shimizu for his useful information. 
N. Yasutake is grateful to T. Inakura for the fruitful discussions.
This work was supported by JSPS KAKENHI Grant Numbers 20K03951, 20H04742.
\end{acknowledgments}

\appendix

\section{Mean weighed method based on independent Gaussian priors}

Here, we introduce how to estimate one unknown value $X$ from some independent measurements.  
In this paper, the kernels of the the priors for the measurements are assumed to be Gaussian types. 
Then, the probability of $i-$th measurement with the physical value $x_i$ is
\begin{eqnarray}
P(x_i)\varpropto\frac{1}{\sigma_i}\exp[-(x_i-X)^2/(2\sigma_i^2)],
\end{eqnarray}
where $\sigma_i$ is each dispersion.

Each measurement is assumed to be independent, therefore the probability $P(x_A,x_B)$ to obtain two measurements $x_A$ and $x_B$, for example, can be expressed as the product of each probability,
\bea
P(x_A,x_B)=P(x_A)P(x_B).
\eea

According to the maximum likelihood principle, 
we obtain the best estimation $X_{\rm best}$ for the unknown quantity, by taking the maximum value of $P(x_A,x_B)$ respect to $X$, as 
\begin{eqnarray}
X_{best}=\frac{w_Ax_A+w_Bx_B}{w_A+w_B},
\end{eqnarray}
where the weights $w_A$ and $w_B$ are given by
\begin{eqnarray}
w_A=1/\sigma_A^2,~~~~~w_B=1/{\sigma_B^2}.
\end{eqnarray}

We can extend the above discussion to general cases with $N$-samples, then we obtain 
\begin{eqnarray}
X_{best}= \sum\limits_{i=1}^N w_ix_i~  \Big/ ~ \sum\limits_{i=1}^N w_i,
\end{eqnarray}
where the weights $w_i$ are related to the individual error distributions in a similar way,
\begin{eqnarray}
w_i=1/\sigma_i^2 ~~~  (i=1,2,...,N).
\end{eqnarray}
The standard deviation from the weighted mean, $\sigma$, is reobtained as 
\begin{eqnarray}
\sigma^2=\frac{\sum\limits_{i=1}^N w_i^2  (x_i-X_{\rm best})\sigma_i^2}{\Big(\sum\limits_{i=1}^N w_i \Big)^2}
=\frac{1}{\sum\limits_{i=1}^N w_i }
\end{eqnarray}

The true value $X$ is expressed in the form,
\begin{eqnarray}
X=X_{\rm best} \pm \sigma.
\end{eqnarray}


\onecolumngrid

\squeezetable
\begin{table}[h]
\caption{Properties of 206 EoSs (1). The symbol $^\ddagger$ is our results, while 
$^\dagger$  denotes the results of self-consistent calculations.   
}
\label{table-I}
\begin{tabular}{c|ccccccc|cc}
\hline
 & m*/m & K &J & L & Ksym & Rskin-208 & Rskin-48 & Refs. \cr
\hline	      
APR, E0019& & 266.000  & 32.600  & 57.600  &  & 0.160   & 0.160$^\dagger$  
& \cite{Akmal:1998cf,Brown:2000pd, Ishizuka:2014jsa}   & \cr
BHF-1 &  & 195.500  & 34.300  & 66.550  & -31.300  & 0.200$^\dagger$  & 0.183$^\dagger$  & 
\cite{Ducoin:2010as}  & \cr
BSk14 & 0.800  & 239.380  & 30.000  & 43.910  & -152.030  & 0.164$^\dagger$  & 0.162$^\dagger$  & 
\cite{Ducoin:2010as,Dutra:2012mb}  &\cr 
BSk16 & 0.800  & 241.730  & 30.000  & 34.870  & -187.390  & 0.149$^\dagger$  & 0.154$^\dagger$  & 
\cite{Ducoin:2010as,Dutra:2012mb} & \cr
BSk17 & 0.800  & 241.740  & 30.000  & 36.280  & -181.860  & 0.151$^\dagger$  & 0.156$^\dagger$  & 
\cite{Ducoin:2010as,Dutra:2012mb}  & \cr
BSk20 & 0.800  & 241.400  & 30.000  & 37.400  & -136.500  & 0.153$^\dagger$  & 0.157$^\dagger$  &  \cite{Fortin:2016hny,Dutra:2012mb}  & \cr
BSk21 & 0.800  & 245.800  & 30.000  & 46.600  & -37.200  & 0.168$^\dagger$  & 0.165$^\dagger$  &  \cite{Fortin:2016hny,Dutra:2012mb}  & \cr
BSk22 &  & 245.900  & 32.000  & 68.500  & 13.000  & 0.204$^\dagger$  & 0.185$^\dagger$  &  
\cite{Fortin:2016hny}  & \cr
BSk23 &  & 245.700  & 31.000  & 57.800  & -11.300  & 0.186$^\dagger$  & 0.175$^\dagger$  & 
\cite{Fortin:2016hny,Ishizuka:2014jsa}   & \cr
BSk24 &  & 245.500  & 30.000  & 46.400  & -37.600  & 0.168$^\dagger$  & 0.165$^\dagger$  & 
\cite{Fortin:2016hny}   & \cr
BSk25 &  & 236.000  & 29.000  & 36.900  & -28.500  & 0.152$^\dagger$  & 0.156$^\dagger$   & 
\cite{Fortin:2016hny}   & \cr
BSk26 &  & 240.800  & 30.000  & 37.500  & -135.600  & 0.153$^\dagger$  & 0.157$^\dagger$  & 
\cite{Fortin:2016hny}   & \cr
BSR2 &  & 239.900  & 31.500  & 62.000  & -3.100  & 0.193$^\dagger$  & 0.179$^\dagger$  & 
\cite{Fortin:2016hny}   & \cr
BSR6 &  & 235.800  & 35.600  & 85.700  & -49.600  & 0.231$^\dagger$  & 0.200$^\dagger$  & 
\cite{Fortin:2016hny}   & \cr
D1 &  & 229.400  & 30.700  & 18.360  & -274.600  & 0.122$^\dagger$  & 0.139$^\dagger$  & 
\cite{Gonzalez-Boquera:2017uep}   & \cr
D1AS &  & 229.400  & 31.300  & 66.550  & -89.100  & 0.200$^\dagger$  & 0.183$^\dagger$  & 
\cite{Gonzalez-Boquera:2017uep}  & \cr
D1M & 0.746$^\ddagger$  & 224.958$^\ddagger$  & 28.552$^\ddagger$  & 24.966$^\ddagger$  & -133.692$^\ddagger$  & 0.113$^\ddagger$  & 0.147$^\ddagger$  & 
\cite{Gonzalez-Boquera:2017rzy}  & \cr
D1M* & 0.746$^\ddagger$  & 225.365$^\ddagger$  & 30.249$^\ddagger$  & 43.311$^\ddagger$  & -47.793$^\ddagger$  & 0.134$^\ddagger$  & 0.158$^\ddagger$  & 
\cite{Gonzalez-Boquera:2017rzy}   & \cr
D1MK & 0.746$^\ddagger$  & 225.400$^\ddagger$  & 33.000$^\ddagger$  & 55.000$^\ddagger$  & -37.275$^\ddagger$  & 0.158$^\ddagger$  & 0.171$^\ddagger$  & TW  & \cr
D1N & 0.748$^\ddagger$  & 225.525$^\ddagger$  & 29.594$^\ddagger$  & 33.665$^\ddagger$  & -168.750$^\ddagger$  & 0.144$^\ddagger$  & 0.171$^\ddagger$  & 
\cite{Gonzalez-Boquera:2017rzy}  & \cr
D1P & 0.672$^\ddagger$  & 250.860$^\ddagger$  & 32.418$^\ddagger$  & 49.827$^\ddagger$  & -157.419$^\ddagger$  & 0.179$^\ddagger$  & 0.157$^\ddagger$  & 
\cite{D1P-99,Gonzalez-Boquera:2017uep}  & \cr
D1PK & 0.700$^\ddagger$  & 260.000$^\ddagger$  & 33.000$^\ddagger$  & 55.000$^\ddagger$  & -150.000$^\ddagger$  & 0.182$^\ddagger$  & 0.181$^\ddagger$  & TW  & \cr
D1S & 0.697$^\ddagger$  & 202.856$^\ddagger$  & 31.125$^\ddagger$  & 22.558$^\ddagger$  & -241.797$^\ddagger$  & 0.137$^\ddagger$  & 0.159$^\ddagger$  & 
\cite{Gonzalez-Boquera:2017rzy,Inakura:2015cla}  & \cr
D2 & 0.738  & 209.300  & 31.130  & 44.850  &  & 0.165$^\dagger$  & 0.163$^\dagger$  & 
\cite{Gonzalez-Boquera:2017rzy}  & \cr
D250 &  & 249.900  & 31.570  & 24.820  & -289.400  & 0.133$^\dagger$  & 0.145$^\dagger$  & 
\cite{Gonzalez-Boquera:2017uep}   & \cr
D260 &  & 259.500  & 30.110  & 17.570  & -298.700  & 0.121$^\dagger$  & 0.139$^\dagger$  & 
\cite{Gonzalez-Boquera:2017uep}   & \cr
D280 &  & 285.200  & 33.140  & 46.530  & -211.900  & 0.168$^\dagger$  & 0.165$^\dagger$  & 
\cite{Gonzalez-Boquera:2017uep}   & \cr
D300 &  & 299.100  & 31.220  & 25.840  & -315.100  & 0.135$^\dagger$  & 0.146$^\dagger$  & 
\cite{Gonzalez-Boquera:2017uep}   & \cr
DD &  & 241.000  & 31.700  & 56.000  & -95.000  & 0.183$^\dagger$  & 0.173$^\dagger$  &  
\cite{Chen:2010qx}  & \cr
DD-F &  & 223.000  & 31.600  & 56.000  & -140.000  & 0.183$^\dagger$  & 0.173$^\dagger$  & 
\cite{Chen:2010qx}  & \cr
DD-ME1 &  & 245.000  & 33.100  & 55.000  & -101.000  & 0.203$^\dagger$  & 0.193$^\dagger$  & 
\cite{Chen:2010qx,Zhao:2016ujh,Ducoin:2010as,Ishizuka:2014jsa}  & \cr
DD-ME2 &  & 251.000  & 32.300  & 51.240  & -87.000  & 0.203$^\dagger$  & 0.187$^\dagger$  & 
\cite{Chen:2010qx,Zhao:2016ujh,Ducoin:2010as,Fortin:2016hny}  & \cr
DD-PC1 &  &  &  & 67.799  &  & 0.203$^\dagger$  & 0.195$^\dagger$  & 
\cite{Zhao:2016ujh,Ducoin:2010as}  & \cr
Ducoin &  & 240.200  & 32.760  & 55.300  & -124.700  & 0.182$^\dagger$  & 0.173$^\dagger$  & 
\cite{Ducoin:2010as}   & \cr
E0008(TMA) &  & 318.000  & 30.660  & 90.140  &  & 0.239$^\dagger$  & 0.204$^\dagger$  & 
\cite{Ishizuka:2014jsa}  & \cr
E0009 &  & 280.000  & 32.500  & 88.700  &  & 0.236$^\dagger$  & 0.203$^\dagger$  & 
\cite{Ishizuka:2014jsa,Oertel:2016bki}  & \cr
E0015 &  & 216.700  & 30.030  & 45.780  &  & 0.167  & 0.164  & 
\cite{Ishizuka:2014jsa}  & \cr
E0024 &  & 244.500  & 33.100  & 55.000  &  & 0.182$^\dagger$  & 0.172  & 
\cite{Ishizuka:2014jsa}  & \cr
E0025 &  & 211.000  & 31.600  & 107.400  &  & 0.267$^\dagger$  & 0.220$^\dagger$  & 
\cite{Ishizuka:2014jsa}  & \cr
E0036 &  & 281.000  & 36.900  & 110.800  &  & 0.272$^\dagger$  & 0.223$^\dagger$  & 
\cite{Ishizuka:2014jsa}  & \cr
es25 &  & 211.730  & 25.000  & 27.749$^\dagger$  &  & 0.138  & 0.148$^\dagger$  & 
\cite{Steiner:2004fi}   & \cr
es275 &  & 205.330  & 27.500  & 48.549$^\dagger$  &  & 0.171  & 0.167$^\dagger$  & 
\cite{Steiner:2004fi,Fortin:2016hny}  & \cr
es30 &  & 215.360  & 30.000  & 69.603$^\dagger$  &  & 0.205  & 0.186$^\dagger$  & 
\cite{Steiner:2004fi}  & \cr
es325 &  & 212.450  & 32.500  & 81.925$^\dagger$  &  & 0.225  & 0.197$^\dagger$  & 
\cite{Steiner:2004fi}  & \cr
es35 &  & 209.970  & 34.937  & 96.182$^\dagger$  &  & 0.248  & 0.210$^\dagger$  & 
\cite{Steiner:2004fi}  & \cr
FKVW &  & 379.000  & 33.100  & 80.000  & 11.000  & 0.222$^\dagger$  & 0.195$^\dagger$  & 
\cite{Chen:2010qx}   & \cr
FSU &  & 230.000  & 32.590  & 60.500  & -51.300  & 0.210  & 0.188$^\dagger$  & 
\cite{Fattoyev:2013yaa,Ducoin:2010as}  & \cr
FSUgold &  & 229.000  & 32.500  & 60.000  & -52.000  & 0.210  & 0.200  & 
\cite{Chen:2010qx,Oertel:2016bki,Piekarewicz:2007dx}    & \cr
FSUgold2.1 &  & 230.000  & 32.590  & 60.500  &  & 0.191$^\dagger$  & 0.177$^\dagger$  & 
\cite{Ishizuka:2014jsa,Oertel:2016bki}   & \cr
GM1 &  & 299.700  & 32.480  & 93.870  & 17.890  & 0.245$^\dagger$  & 0.207$^\dagger$  & 
\cite{Ducoin:2010as,Fortin:2016hny}   & \cr
GM3 &  & 239.900  & 32.480  & 89.660  & -6.470  & 0.238$^\dagger$  & 0.204$^\dagger$  & 
\cite{Ducoin:2010as}  & \cr
Gs &  & 237.570  & 31.384  & 89.304$^\dagger$  &  & 0.237  & 0.203$^\dagger$  & 
\cite{Steiner:2004fi}  & \cr
GSkI &  & 230.210  & 32.030  & 63.450  & -95.290  & 0.195$^\dagger$  & 0.180$^\dagger$  & 
\cite{Dutra:2012mb}   & \cr
GSkII & 0.790  & 233.400  & 30.490  & 48.630  & -157.830  & 0.171$^\dagger$  & 0.167$^\dagger$  & \cite{Dutra:2012mb}   & \cr
GT2 &  & 228.100  & 33.940  & 5.020  & -445.900  & 0.101$^\dagger$  & 0.127$^\dagger$  & 
\cite{Gonzalez-Boquera:2017uep}  & \cr
G$\sigma$ &  & 237.290  & 31.370  & 94.020  & 13.990  & 0.245$^\dagger$  & 0.208$^\dagger$  & 
\cite{Ducoin:2010as}  & \cr
HA &  & 233.000  & 30.700  & 55.000  & -135.000  & 0.182$^\dagger$  & 0.172$^\dagger$  & 
\cite{Chen:2010qx}  & \cr
HFB-17 &  &  &  & 36.300  &  & 0.151  & 0.155$^\dagger$  & 
\cite{RocaMaza:2011pm}   & \cr
HFB-8 &  &  &  & 14.800  &  & 0.115  & 0.135$^\dagger$  & 
\cite{RocaMaza:2011pm}  & \cr
HS(DD2) &  & 243.000  & 31.700  & 55.000  & -93.200  & 0.182$^\dagger$  & 0.172$^\dagger$  & 
\cite{Oertel:2016bki,Fortin:2016hny}  & \cr
IU-FSU &  & 231.200  & 31.300  & 47.200  & 28.700  & 0.160  & 0.160$^\dagger$  & 
\cite{Fattoyev:2013yaa, Oertel:2016bki}    & \cr
KDE0v1 & 0.740  & 227.540  & 34.580  & 54.690  & -127.120  & 0.181$^\dagger$  & 0.172$^\dagger$  & 
\cite{Dutra:2012mb}   & \cr
KDE0v1-B & 0.790  & 216.000  & 34.900  & 61.000  &  & 0.192  & 0.172  & 
\cite{Brown:2013pwa}  & \cr
KDE0v1-T & 0.810  & 217.000  & 34.600  & 72.000  & -40.000  & 0.200  & 0.178  & 
\cite{Tsang:2019ymt}   & \cr
\hline
\end{tabular}
\end{table} 
\setcounter{table}{0}
\begin{table}[h]
\caption{Properties of 206 EoSs (2).
}
\begin{tabular}{c|ccccccc|cc}
\hline
 & m*/m & K &J & L & Ksym & Rskin-208 & Rskin-48 & Refs. \cr
\hline
LNS & 0.830  & 210.780  & 33.430  & 61.450  & -127.360  & 0.192$^\dagger$  & 0.178$^\dagger$  & 
\cite{Dutra:2012mb,Ducoin:2010as}     & \cr
LS180 &  & 180.000  & 28.600  & 73.800  &  & 0.212$^\dagger$  & 0.189$^\dagger$  & 
\cite{Dutra:2012mb,Ishizuka:2014jsa,Oertel:2016bki}   & \cr
LS220 &  & 220.000  & 28.600  & 73.800  &  & 0.212$^\dagger$  & 0.189$^\dagger$  & 
\cite{Dutra:2012mb,Ishizuka:2014jsa,Oertel:2016bki}   & \cr
LS375 &  & 375.000  & 28.600  & 73.800  &  & 0.212$^\dagger$  & 0.189$^\dagger$  & 
\cite{Dutra:2012mb,Ishizuka:2014jsa,Oertel:2016bki}  & \cr
Ly5 &  & 229.940  & 32.010  & 45.243$^\dagger$  &  & 0.166  & 0.164$^\dagger$  & 
\cite{Steiner:2004fi}   & \cr
M3Y-P6 &  & 239.700  & 32.100  & 44.600  & -165.300  & 0.165$^\dagger$  & 0.163$^\dagger$  & 
\cite{Inakura:2015cla,Lim:2013tqa}  & \cr
M3Y-P7 &  & 254.700  & 31.700  & 51.500  & -127.800  & 0.176$^\dagger$  & 0.169$^\dagger$  & 
\cite{Inakura:2015cla,Fortin:2016hny}  & \cr
MSk3 & 1.000  & 233.250  & 28.000  & 7.040  & -283.520  & 0.111  & 0.128  &  
\cite{Dutra:2012mb,Zhao:2016ujh}   & \cr
MSk6 & 1.050  & 231.170  & 28.000  & 9.630  & -274.330  & 0.118  & 0.130  &  
\cite{Dutra:2012mb,Zhao:2016ujh}   & \cr
MSk7 & 1.050  & 385.360  & 27.950  & 9.400  & -274.630  & 0.116  & 0.136$^\dagger$  & 
\cite{RocaMaza:2011pm,Dutra:2012mb}   & \cr
MSL0 & 0.800  & 230.000  & 30.000  & 60.000  & -99.330  & 0.180  & 0.171$^\dagger$  & 
\cite{Chen:2010qx,Dutra:2012mb,Wang:2014rva}  & \cr
NL1 &  & 212.000  & 43.500  & 140.000  & 143.000  & 0.319  & 0.247  & 
\cite{Chen:2010qx,Zhao:2016ujh}   & \cr
NL2 &  & 401.000  & 44.000  & 130.000  & 20.000  & 0.304  & 0.243  & 
\cite{Chen:2010qx,Zhao:2016ujh}   & \cr
NL3 &  & 271.000  & 37.300  & 118.000  & 100.000  & 0.280  & 0.230  & 
\cite{Chen:2010qx,Fattoyev:2013yaa,Centelles:2010qh,Piekarewicz:2007dx,Ducoin:2010as,Fortin:2016hny} 
 & \cr
NL3* &  &  &  & 119.769$^\dagger$  &  & 0.287  & 0.230  & 
\cite{Zhao:2016ujh}  & \cr
NL3$\omega \rho$ &  & 271.600  & 31.700  & 55.500  & -7.600  & 0.183$^\dagger$  & 0.173$^\dagger$  & \cite{Fortin:2016hny}  & \cr
NL4 &  & 270.350  & 36.239  & 111.649$^\dagger$  &  & 0.273  & 0.223$^\dagger$  & 
\cite{Steiner:2004fi}   & \cr
NL-SH &  & 356.000  & 36.100  & 114.000  & 80.000  & 0.263  & 0.214  & 
\cite{Chen:2010qx,Zhao:2016ujh}   & \cr
NL$\rho$ &  & 240.000  & 30.300  & 85.000  & 3.000  & 0.230$^\dagger$  & 0.199$^\dagger$  & 
\cite{Chen:2010qx}  & \cr
NL$\omega \rho$(025) &  & 270.700  & 32.350  & 61.050  & -34.360  & 0.192$^\dagger$  & 0.178$^\dagger$  & \cite{Ducoin:2010as}  & \cr
NRAPR & 0.690  & 225.700  & 32.787  & 59.630  & -123.320  & 0.190  & 0.177$^\dagger$  & 
\cite{Steiner:2004fi,Dutra:2012mb}   & \cr
NRAPR-B & 0.850  & 225.000  & 35.100  & 61.000  &  & 0.193  & 0.178  & 
\cite{Brown:2013pwa}  & \cr
NRAPR-T & 0.730  & 221.000  & 34.100  & 70.000  & -46.000  & 0.195  & 0.181  & 
\cite{Tsang:2019ymt}  & \cr
PC-F1 &  & 255.000  & 37.800  & 117.000  & 75.000  & 0.269  & 0.225  & 
\cite{Chen:2010qx,Zhao:2016ujh}  & \cr
PC-F2 &  & 256.000  & 37.600  & 116.000  & 65.000  & 0.281$^\dagger$  & 0.227$^\dagger$  & 
\cite{Chen:2010qx,Zhao:2016ujh,Wang:2014rva}  & \cr
PC-F3 &  & 256.000  & 38.300  & 119.000  & 74.000  & 0.285$^\dagger$  & 0.230$^\dagger$  & 
\cite{Chen:2010qx,Zhao:2016ujh} & \cr
PC-F4 &  & 255.000  & 37.700  & 119.000  & 98.000  & 0.285$^\dagger$  & 0.230$^\dagger$  & 
\cite{Chen:2010qx}  & \cr
PC-LA &  & 263.000  & 37.200  & 108.000  & -61.000  & 0.268$^\dagger$  & 0.220$^\dagger$  & 
\cite{Chen:2010qx}  & \cr
PC-PK1 &  &  &  & 101.478  &  & 0.257  & 0.220  & 
\cite{Zhao:2016ujh}  & \cr
PK1 &  & 282.000  & 37.600  & 116.000  & 55.000  & 0.277  & 0.223  & 
\cite{Chen:2010qx,Zhao:2016ujh}   & \cr
PKDD &  & 263.000  & 36.900  & 90.000  & -80.000  & 0.253  & 0.214  & 
\cite{Chen:2010qx,Zhao:2016ujh}   & \cr
RAPR &  & 276.700  & 33.987  & 66.958$^\dagger$  &  & 0.201  & 0.183$^\dagger$  & 
\cite{Steiner:2004fi}   & \cr
RATP &  & 239.580  & 29.260  & 32.390  & -191.250  & 0.145$^\dagger$  & 0.152$^\dagger$  & 
\cite{Ducoin:2010as}  & \cr
rDD-ME2 &  &  &  & 51.300  &  & 0.193  & 0.179$^\dagger$  & 
\cite{RocaMaza:2011pm}   & \cr
rFSUGold &  &  &  & 60.500  &  & 0.207  & 0.186$^\dagger$  & 
\cite{RocaMaza:2011pm,Fortin:2016hny}  & \cr
rG2 &  &  &  & 100.700  &  & 0.257  & 0.214$^\dagger$  & 
\cite{RocaMaza:2011pm}  & \cr
rNL1 &  &  &  & 140.100  &  & 0.321  & 0.250$^\dagger$  & 
\cite{RocaMaza:2011pm}  & \cr
rNL3 &  &  &  & 118.500  &  & 0.280  & 0.227$^\dagger$  & 
\cite{RocaMaza:2011pm}   & \cr
rNL3* &  &  &  & 122.600  &  & 0.288  & 0.232$^\dagger$  & \cite{RocaMaza:2011pm}  & \cr
rNLC &  &  &  & 108.000  &  & 0.263  & 0.218$^\dagger$  & 
\cite{RocaMaza:2011pm}  & \cr
rNL-RA1 &  &  &  & 115.400  &  & 0.274  & 0.224$^\dagger$  & 
\cite{RocaMaza:2011pm}  & \cr
rNL-SH &  &  &  & 113.600  &  & 0.266  & 0.219$^\dagger$  & 
\cite{RocaMaza:2011pm}  & \cr
rNL-Z &  &  &  & 133.300  &  & 0.307  & 0.242$^\dagger$  & 
\cite{RocaMaza:2011pm}  & \cr
Rs &  & 237.660  & 30.593  & 80.096$^\dagger$  & -9.100  & 0.222  & 0.195$^\dagger$  & 
\cite{Steiner:2004fi,Fortin:2016hny}  & \cr
rTM1 &  &  &  & 110.800  &  & 0.271  & 0.222$^\dagger$  & 
\cite{RocaMaza:2011pm}  & \cr
R$\sigma$&  & 237.410  & 30.580  & 85.700  & -9.130  & 0.231$^\dagger$  & 0.200$^\dagger$  & 
\cite{Ducoin:2010as}  & \cr
S271 &  & 271.000  & 35.927  & 97.541$^\dagger$  &  & 0.251  & 0.211$^\dagger$  & 
\cite{Steiner:2004fi}   & \cr
SFHo &  & 245.000  & 31.600  & 47.100  &  & 0.169$^\dagger $ & 0.165$^\dagger$  & 
\cite{Oertel:2016bki}  & \cr
SFHx &  & 239.000  & 28.700  & 23.200  &  & 0.130$^\dagger$  & 0.144$^\dagger$  & 
\cite{Oertel:2016bki}  & \cr
SGI & 0.610  & 262.000  & 28.300  & 63.900  & -51.990  & 0.196$^\dagger$  & 0.180$^\dagger$  & 
\cite{Steiner:2004fi,Lim:2013tqa,Dutra:2012mb}  & \cr
SGII & 0.790  & 214.700  & 26.830  & 37.620  & -145.920  & 0.136  & 0.147$^\dagger$  & 
\cite{Ducoin:2010as,RocaMaza:2011pm,Inakura:2015cla,Dutra:2012mb} & \cr
SII & 0.580  & 341.400  & 34.160  & 50.020  & -265.720  & 0.196  & 0.177  &   
\cite{Zhao:2016ujh}  & \cr
SIII & 0.760  & 355.37 & 28.160  & 9.910  & -393.730  & 0.137  & 0.125  & 
\cite{Dutra:2012mb,Zhao:2016ujh}   & \cr
Sk$\chi$m &  & 230.400  & 30.940  & 45.600  &  & 0.167  & 0.164$^\dagger$  & 
\cite{Zhang:2017hvh,Dutra:2012mb}   & \cr
SK255 &  & 254.960  & 37.400  & 95.000  & -58.300  & 0.247$^\dagger$  & 0.208$^\dagger$  & 
\cite{Fortin:2016hny}  & \cr
SK272 &  & 271.550  & 37.400  & 91.700  & -67.800  & 0.241$^\dagger$  & 0.205$^\dagger$  & 
\cite{Fortin:2016hny}  & \cr
Ska & 0.610  & 263.160  & 32.910  & 74.620  & -78.460  & 0.214$^\dagger$  & 0.190$^\dagger$  & 
\cite{RocaMaza:2011pm,Dutra:2012mb,Fortin:2016hny,Oertel:2016bki,Inakura:2015cla,Wang:2014rva}  & \cr
Ska25-B & 0.990  & 219.000  & 32.500  & 51.000  &  & 0.176  & 0.170  & 
\cite{Brown:2013pwa}  & \cr
Ska25s20 & 0.980  & 220.750  & 33.780  & 63.810  & -118.220  & 0.196$^\dagger$  & 0.180$^\dagger$  & \cite{Dutra:2012mb}  & \cr
Ska25-T & 0.980  & 220.000  & 31.900  & 59.000  & -59.000  & 0.183  & 0.176  & 
\cite{Tsang:2019ymt}  & \cr
Ska35-B & 1.000  & 244.000  & 32.800  & 54.000  &  & 0.180  & 0.172  & 
\cite{Brown:2013pwa}  & \cr
Ska35s20 & 1.000  & 240.270  & 33.570  & 64.830  & -120.320  & 0.198$^\dagger$  & 0.181$^\dagger$  & \cite{Dutra:2012mb,Wang:2014rva}  & \cr
Ska35-T & 0.990  & 238.000  & 32.000  & 58.000  & -84.000  & 0.184  & 0.177  & 
\cite{Tsang:2019ymt,Dutra:2012mb,Ducoin:2010as}   & \cr
\hline
\end{tabular}
\end{table} 

\setcounter{table}{0}
\begin{table}[h]
\caption{Properties of 206 EoSs (3).
}
\begin{tabular}{c|ccccccc|cc}
\hline
 & m*/m & K &J & L & Ksym & Rskin-208 & Rskin-48 & Refs. \cr
\hline
SKb & 0.610  & 263.000  & 33.880  & 47.600  & -78.500  & 0.170$^\dagger$  & 0.166$^\dagger$  & 
\cite{Fortin:2016hny,Dutra:2012mb}  & \cr
SkI1 & 0.690  & 242.750  & 37.530  & 161.050  & 234.670  & 0.353$^\dagger$  & 0.268$^\dagger$  & 
\cite{Wang:2014rva,Dutra:2012mb}  & \cr
SkI2 & 0.680  & 240.700  & 33.400  & 104.300  & 70.600  & 0.262$^\dagger$  & 0.217$^\dagger$  &  \cite{Ducoin:2010as,Fortin:2016hny,Dutra:2012mb,Inakura:2015cla}   & \cr
SkI3 & 0.580  & 258.000  & 34.800  & 100.500  & 72.900  & 0.255$^\dagger$  & 0.213$^\dagger$  & 
\cite{Ducoin:2010as,Fortin:2016hny,Dutra:2012mb,Inakura:2015cla} & \cr
SkI4 & 0.650  & 247.700  & 29.500  & 60.400  & -40.600  & 0.191$^\dagger$  & 0.177$^\dagger$  & 
\cite{Ducoin:2010as,Fortin:2016hny,Dutra:2012mb,Inakura:2015cla,Lim:2013tqa}  & \cr
SkI5 & 0.580  & 255.800  & 36.697  & 129.300  & 159.500  & 0.272  & 0.214  & 
\cite{Steiner:2004fi,Ducoin:2010as,Fortin:2016hny,Dutra:2012mb,Inakura:2015cla}   & \cr
SkI6 & 0.640  & 248.650  & 30.090  & 59.700  & -47.270  & 0.189$^\dagger$  & 0.177$^\dagger$  & 
\cite{Ducoin:2010as,Fortin:2016hny,Dutra:2012mb}   & \cr
SkM* & 0.790  & 216.610  & 30.030  & 45.780  & -155.940  & 0.170  & 0.155  & 
\cite{RocaMaza:2011pm,Dutra:2012mb,Inakura:2015cla,Zhao:2016ujh}  & \cr
SkM*-B & 0.780  & 218.000  & 34.200  & 58.000  &  & 0.187  & 0.175  & 
\cite{Brown:2013pwa}  & \cr
SkM*-T & 0.790  & 219.000  & 33.700  & 65.000  & -65.000  & 0.187  & 0.179  & 
\cite{Tsang:2019ymt} & \cr
SkMP & 0.650  & 230.930  & 29.890  & 70.310  & -49.820  & 0.197  & 0.167  & 
\cite{RocaMaza:2011pm,Ducoin:2010as,Steiner:2004fi,Fortin:2016hny}  & \cr
SkO & 0.900  & 223.390  & 31.970  & 79.140  & -43.170  & 0.221$^\dagger$  & 0.194$^\dagger$  & 
\cite{Ducoin:2010as,Dutra:2012mb}   & \cr
SKOp & 0.900  & 222.360  & 31.950  & 68.940  & -78.820  & 0.204$^\dagger$  & 0.185$^\dagger$ & \cite{Dutra:2012mb,Fortin:2016hny}  & \cr
SKP & 1.000  & 200.970  & 30.000  & 19.680  & -266.600  & 0.144  & 0.144  &  
\cite{Dutra:2012mb,Zhao:2016ujh}   & \cr
SKRA & 0.750  & 216.980  & 31.320  & 53.040  & -139.280  & 0.179$^\dagger$  & 0.171$^\dagger$  & 
\cite{Dutra:2012mb}   & \cr
SKRA-B & 0.790  & 212.000  & 33.700  & 55.000  &  & 0.181  & 0.172  & 
\cite{Brown:2013pwa,Dutra:2012mb}  & \cr
SKRA-T & 0.800  & 213.000  & 33.400  & 65.000  & -55.000  & 0.190  & 0.179  & 
\cite{Tsang:2019ymt,Dutra:2012mb}   & \cr
Sk-Rs &  &  &  & 85.700  &  & 0.215  & 0.191$^\dagger$  & 
\cite{RocaMaza:2011pm}  & \cr
SkSM* &  &  &  & 65.500  &  & 0.197  & 0.181$^\dagger$  & 
\cite{RocaMaza:2011pm}  & \cr
SkT1 & 1.000  & 236.160  & 32.020  & 56.180  & -134.830  & 0.184$^\dagger$  & 0.173$^\dagger$  &   
\cite{Dutra:2012mb}   & \cr
SkT1-B & 0.970  & 242.000  & 33.300  & 56.000  &  & 0.183  & 0.172  & 
\cite{Brown:2013pwa}  & \cr
SkT1-T & 0.970  & 238.000  & 32.600  & 63.000  & -70.000  & 0.190  & 0.179  & 
\cite{Tsang:2019ymt,Dutra:2012mb}   & \cr
SkT2 & 1.000  & 235.730  & 32.000  & 56.160  & -134.670  & 0.184$^\dagger$  & 0.173$^\dagger$  &   
\cite{Dutra:2012mb}   & \cr
SkT2-B & 0.970  & 242.000  & 33.500  & 58.000  &  & 0.186  & 0.174  & 
\cite{Brown:2013pwa}  & \cr
SkT2-T & 0.960  & 238.000  & 32.600  & 62.000  & -75.000  & 0.188  & 0.178  & 
\cite{Tsang:2019ymt,Dutra:2012mb}   & \cr
SkT3 & 1.000  & 235.740  & 31.500  & 55.310  & -132.050  & 0.182$^\dagger$  & 0.173$^\dagger$  &  
\cite{Dutra:2012mb}   & \cr
SkT3-B & 0.980  & 241.000  & 32.700  & 53.000  &  & 0.179  & 0.172  & 
\cite{Brown:2013pwa}  & \cr
SkT3-T & 0.970  & 236.000  & 31.900  & 58.000  & -80.000  & 0.183  & 0.178  & 
\cite{Tsang:2019ymt}  & \cr
Sk-T4 & 1.000  & 235.560  & 35.457  & 94.100  & -24.500  & 0.253  & 0.212$^\dagger $ & 
\cite{Steiner:2004fi,Inakura:2015cla,RocaMaza:2011pm}  & \cr
Sk-T6 & 1.000  & 235.950  & 29.970  & 30.900  & -211.530  & 0.151  & 0.155$^\dagger $ & 
\cite{RocaMaza:2011pm,Dutra:2012mb}   & \cr
Skxs20 & 0.960  & 201.950  & 35.500  & 67.060  & -122.310  & 0.201$^\dagger$  & 0.183$^\dagger$  & \cite{Dutra:2012mb}   & \cr
Skz2 & 0.700  & 230.070  & 32.010  & 16.810  & -259.660  & 0.120$^\dagger$  & 0.138$^\dagger$  &  \cite{Dutra:2012mb,Wang:2014rva}   & \cr
Skz4 & 0.700  & 230.080  & 32.010  & 5.750  & -240.860  & 0.102$^\dagger$  & 0.128$^\dagger$  &   \cite{Dutra:2012mb,Wang:2014rva}   & \cr
SLy0 & 0.700  & 229.670  & 31.982  & 44.873$^\dagger$  & -116.230  & 0.165  & 0.163$^\dagger$  & 
\cite{Steiner:2004fi,Dutra:2012mb}   & \cr
SLy10 & 0.680  & 229.740  & 31.980  & 38.740  & -142.190  & 0.155$^\dagger$  & 0.158$^\dagger$  & 
\cite{Ducoin:2010as,Dutra:2012mb}   & \cr
Sly2 & 0.700  & 229.920  & 32.000  & 47.460  & -115.130  & 0.170$^\dagger$  & 0.166$^\dagger$ & \cite{Dutra:2012mb,Fortin:2016hny}   & \cr
SLy230a & 0.700  & 229.890  & 31.980  & 44.310  & -98.210  & 0.155  & 0.158$^\dagger$  & 
\cite{Steiner:2004fi,Ducoin:2010as,Fortin:2016hny,Dutra:2012mb}    & \cr
Sly230b & 0.690  & 229.960  & 32.010  & 45.960  & -119.720  & 0.167$^\dagger$  & 0.164$^\dagger$  & \cite{Ducoin:2010as,Dutra:2012mb}   & \cr
SLy4 & 0.690  & 229.900  & 32.000  & 45.900  & -119.700  & 0.162  & 0.152  & 
\cite{RocaMaza:2011pm,Lim:2013tqa,Inakura:2015cla,Gonzalez-Boquera:2017rzy,Fortin:2016hny,Zhao:2016ujh}   & \cr
SLy4-B & 0.700  & 224.000  & 34.100  & 56.000  &  & 0.184  & 0.174  & 
\cite{Brown:2013pwa}  & \cr
SLy4-T & 0.760  & 222.000  & 33.600  & 66.000  & -55.000  & 0.191  & 0.179  & 
\cite{Tsang:2019ymt,Dutra:2012mb}   & \cr
SLy5 & 0.700  & 229.920  & 32.010  & 48.150  & -112.760  & 0.162  & 0.160  &
\cite{Dutra:2012mb,Zhao:2016ujh}   & \cr
SLy6 & 0.690  & 229.860  & 31.960  & 47.450  & -112.710  & 0.161  & 0.152   &
\cite{Dutra:2012mb,Zhao:2016ujh}   & \cr
Sly9 & 0.670  & 229.840  & 31.980  & 54.860  & -81.420  & 0.182$^\dagger$  & 0.172$^\dagger$  & \cite{Dutra:2012mb,Fortin:2016hny}   & \cr
SQMC650 & 0.780  & 218.110  & 33.650  & 52.920  & -173.150  & 0.178$^\dagger$  & 0.171$^\dagger$  & \cite{Dutra:2012mb}   & \cr
SQMC700 & 0.760  & 222.200  & 33.470  & 59.060  & -140.840  & 0.188$^\dagger$  & 0.176$^\dagger$  & \cite{Dutra:2012mb}   & \cr
SQMC750-B & 0.710  & 228.000  & 34.800  & 59.000  &  & 0.190  & 0.176  & 
\cite{Brown:2013pwa}  & \cr
SQMC750-T & 0.750  & 223.000  & 33.900  & 68.000  & -50.000  & 0.194  & 0.180  & 
\cite{Tsang:2019ymt,Dutra:2012mb}   & \cr
SR1 & 0.900  & 202.150  & 29.000  & 41.245$^\dagger$  &  & 0.160  & 0.160$^\dagger$  & 
\cite{Steiner:2004fi}  & \cr
SR2 &  & 224.640  & 30.071  & 49.130$^\dagger$  &  & 0.172  & 0.167$^\dagger$  & 
\cite{Steiner:2004fi}   & \cr
SR3 &  & 222.550  & 29.001  & 48.308$^\dagger$  &  & 0.171  & 0.166$^\dagger$  & 
\cite{Steiner:2004fi}   & \cr
SV & 0.380  & 306.000  & 32.800  & 96.100  & 24.190  & 0.230  & 0.196  & 
\cite{Lim:2013tqa,Ducoin:2010as,Zhao:2016ujh,Dutra:2012mb}    & \cr
SV-bas & 0.900  & 221.760  & 30.000  & 32.000  & -156.570  & 0.155  & 0.158$^\dagger$  & 
\cite{Reinhard:2016sce,Dutra:2012mb}  & \cr
SV-K218 & 0.900  & 218.230  & 30.000  & 35.000  & -206.870  & 0.161  & 0.161$^\dagger$  & 
\cite{Reinhard:2016sce,Dutra:2012mb}  & \cr
SV-K226 & 0.900  & 225.820  & 30.000  & 34.000  & -211.920  & 0.159  & 0.160$^\dagger$  & 
\cite{Reinhard:2016sce,Dutra:2012mb}   & \cr
SV-K241 & 0.900  & 241.070  & 30.000  & 31.000  & -230.770  & 0.151  & 0.155$^\dagger$  & 
\cite{Reinhard:2016sce,Dutra:2012mb}  & \cr
SV-kap00 & 0.900  & 233.440  & 30.000  & 40.000  & -161.780  & 0.158  & 0.159$^\dagger$  & 
\cite{Reinhard:2016sce,Dutra:2012mb}  & \cr
SV-kap20 & 0.900  & 233.440  & 30.000  & 36.000  & -193.190  & 0.155  & 0.158$^\dagger$  & 
\cite{Reinhard:2016sce,Dutra:2012mb}  & \cr
SV-kap60 & 0.900  & 233.450  & 30.000  & 29.000  & -249.750  & 0.154  & 0.157$^\dagger$  & 
\cite{Reinhard:2016sce,Dutra:2012mb}   & \cr
SV-L25 & 0.900  &  & 30.000  & 25.000  &  & 0.143  & 0.151$^\dagger$  & 
\cite{Reinhard:2016sce}   & \cr
SV-L32 & 0.900  &  & 30.000  & 32.000  &  & 0.154  & 0.157$^\dagger$  & 
\cite{Reinhard:2016sce}  & \cr
SV-L40 & 0.900  & 233.3 & 30.000  & 40.000  &  & 0.166  & 0.164$^\dagger$  & 
\cite{Reinhard:2016sce}  & \cr
SV-L47 & 0.900  & 233.4 & 30.000  & 47.000  &  & 0.177  & 0.170$^\dagger$  & 
\cite{Reinhard:2016sce}  & \cr
\hline
\end{tabular}
\end{table}

\setcounter{table}{0}
\begin{table}[h]
\caption{Properties of 206 EoSs (4).
}
\begin{tabular}{c|ccccccc|cc}
\hline
 & m*/m & K &J & L & Ksym & Rskin-208 & Rskin-48 & Refs. \cr
\hline
SV-mas07 & 0.700  & 233.540  & 30.000  & 52.000  & -98.770  & 0.152  & 0.156$^\dagger$  & 
\cite{Reinhard:2016sce,Dutra:2012mb}  & \cr
SV-mas08 & 0.800  & 233.130  & 30.000  & 40.000  & -172.380  & 0.160  & 0.160$^\dagger$  & 
\cite{Reinhard:2016sce,Dutra:2012mb,Wang:2014rva}  & \cr
SV-mas10 & 1.000  & 234.330  & 30.000  & 28.000  & -252.500  & 0.152  & 0.156$^\dagger$  & 
\cite{Reinhard:2016sce,Dutra:2012mb}  & \cr
SV-sym28 & 0.900  & 240.860  & 28.000  & 7.000  & -305.940  & 0.117  & 0.136$^\dagger$  & 
\cite{Reinhard:2016sce,Dutra:2012mb}  & \cr
SV-sym32 & 0.900  & 233.810  & 32.000  & 57.000  & -148.790  & 0.192  & 0.178$^\dagger$  & 
\cite{Reinhard:2016sce,Dutra:2012mb}  & \cr
SV-sym32-B & 0.910  & 237.000  & 32.300  & 51.000  &  & 0.176  & 0.174  & 
\cite{Brown:2013pwa}  & \cr
SV-sym32-T & 0.910  & 232.000  & 31.500  & 58.000  & -77.000  & 0.181  & 0.179  & 
\cite{Tsang:2019ymt}  & \cr
SV-sym34 & 0.900  & 234.070  & 34.000  & 81.000  & -79.080  & 0.227  & 0.198$^\dagger$  &   
\cite{Reinhard:2016sce,Dutra:2012mb,Wang:2014rva}  & \cr
TFa &  & 245.100  & 35.050  & 82.500  & -68.400  & 0.250  & 0.210$^\dagger$  & 
\cite{Fattoyev:2013yaa} & \cr
TFb &  & 250.100  & 40.070  & 122.500  & 45.800  & 0.300  & 0.238$^\dagger$  & 
\cite{Fattoyev:2013yaa}  & \cr
TFc &  & 260.500  & 43.670  & 135.200  & 51.600  & 0.330  & 0.255$^\dagger$  & 
\cite{Fattoyev:2013yaa} & \cr
TM1 &  & 281.000  & 36.900  & 110.800  & 33.550  & 0.272$^\dagger$  & 0.223$^\dagger$  & 
\cite{Ishizuka:2014jsa,Ducoin:2010as,Fortin:2016hny,Chen:2010qx}  & \cr
TW99 &  & 241.000  & 32.800  & 55.000  & -124.000  & 0.196  & 0.186  &  
\cite{Chen:2010qx,Zhao:2016ujh}  &\cr 
UNEDF0 &  & 229.800  & 30.500  & 45.100  & -189.600  & 0.166$^\dagger$  & 0.164$^\dagger$  & 
\cite{Inakura:2015cla,Dutra:2012mb} &\cr 
UNEDF1 &  & 219.800  & 29.000  & 40.000  & -179.400  & 0.158$^\dagger$  & 0.159$^\dagger$  & 
\cite{Inakura:2015cla}  & \cr
Z271 &  & 271.000  & 35.369  & 89.520$^\dagger$  &  & 0.238  & 0.204$^\dagger$  & 
\cite{Steiner:2004fi}   & \cr
\hline
\end{tabular}
\end{table}   


\squeezetable
\begin{table}[h]
\caption{
Parameter sets of D1MK and D1PK. 
}
 \begin{tabular}{c|c|cccc|ccc|c}
\hline
  D1MK &$\mu_i$&$W_i$ & $B_i$ & $H_i$ & $M_i$ &
  $t_0^i$ & $x_0^i$ & $\alpha_i$& $W_0$\cr
\hline
   $i=1$ & 0.5 &  -17242.0144 & 19604.4056 & -20699.9856& 16408.6002&
  1561.7167 & 1 & 1/3  & 115.36 \cr
  $i=2$ & 1.0 & 642.607965& -941.150253& 865.572486& -845.300794&
  0 & -1 & 1 & \cr
\hline
  D1PK &$\mu_i$&$W_i$ & $B_i$ & $H_i$ & $M_i$ &
  $t_0^i$ & $x_0^i$ & $\alpha_i$& $W_0$\cr
\hline
   $i=1$ & 0.90 &  -465.027582 & 155.134492 & -506.775323& 117.749903&
  981.065351& 1 & 1/3  & 130\cr
  $i=2$ & 1.44 &34.6200000& -14.0800000& 70.9500000& -41.3518104&
  534.155654 & -1 & 1 & \cr
\hline
\end{tabular}
\end{table}  

\squeezetable

\begin{table}[h]
\caption{Properties of 47 EoSs satisfying the observational constraint.
}
 \begin{tabular}{c|ccccc|c}
\hline
& J & L & Ksym & R-max & M-max & Refs. \cr
\hline
APR, E0019 & 32.60  & 57.60  &  & 10.0  & 2.20  & \cite{Akmal:1998cf,Brown:2000pd, Ishizuka:2014jsa} \cr
BSk20 & 30.00  & 37.40  & -136.50  & 10.2  & 2.17 & \cite{Fortin:2016hny,Dutra:2012mb} \cr
BSk21 & 30.00  & 46.60  & -37.20  & 11.0  & 2.29 & \cite{Fortin:2016hny,Dutra:2012mb} \cr
BSk22 & 32.00  & 68.50  & 13.00  & 11.2  & 2.27 & \cite{Fortin:2016hny} \cr
BSk23 & 31.00  & 57.80  & -11.30  & 11.2  & 2.28 & \cite{Fortin:2016hny,Ishizuka:2014jsa} \cr
BSk24 & 30.00  & 46.40  & -37.60  & 11.2  & 2.29 & \cite{Fortin:2016hny} \cr
BSk25 & 29.00  & 36.90  & -28.50  & 11.2  & 2.23 & \cite{Fortin:2016hny} \cr
BSk26 & 30.00  & 37.50  & -135.60  & 10.2  & 2.18 & \cite{Fortin:2016hny} \cr
BSR2 & 31.50  & 62.00  & -3.10  & 11.9  & 2.38 & \cite{Fortin:2016hny} \cr
BSR6 & 35.60  & 85.70  & -49.60  & 12.1  & 2.44 & \cite{Fortin:2016hny} \cr
D1AS & 31.30  & 66.55  & -89.10  & 10.0  & 2.00  & \cite{Gonzalez-Boquera:2017uep} \cr
D1M* & 30.25  & 43.15  & -47.79  & 9.6  & 2.00  & \cite{Gonzalez-Boquera:2017rzy,Lourenco:2020qft} \cr
D1PK & 33.00  & 55.00  & -150.00  & 10.0  & 2.1 & TW \cr
D2 & 31.13  & 44.85  &  & 10.2  & 2.1  & \cite{Gonzalez-Boquera:2017rzy,Lourenco:2020qft} \cr
DD-ME1 & 33.10  & 55.00  &  & 11.9  & 2.47  & \cite{Chen:2010qx,Zhao:2016ujh,Ducoin:2010as,Ishizuka:2014jsa} \cr
DD-ME2 & 32.30  & 51.20  & -87.00  & 12.2  & 2.48 & \cite{Chen:2010qx,Zhao:2016ujh,Ducoin:2010as,Fortin:2016hny} \cr
E0008(TMA) & 30.66  & 90.14  &  & 12.4  & 2.04  & \cite{Ishizuka:2014jsa} \cr
FSUgold2.1 & 32.60  & 60.50  &  & 12.1  & 2.13  & \cite{Ishizuka:2014jsa,Oertel:2016bki} \cr
GM1 & 32.50  & 94.40  & 17.89  & 12.0  & 2.36 & \cite{Ducoin:2010as,Fortin:2016hny} \cr
HS(DD2) & 31.70  & 55.00  & -93.20  & 11.8  & 2.42 & \cite{Oertel:2016bki,Fortin:2016hny} \cr
LS220 & 28.60  & 73.80  &  & 10.6  & 2.05  & \cite{Dutra:2012mb,Ishizuka:2014jsa,Oertel:2016bki} \cr
LS375 & 28.60  & 73.80  &  & 12.6  & 2.72  & \cite{Dutra:2012mb,Ishizuka:2014jsa,Oertel:2016bki} \cr
NL3 & 37.30  & 118.00  & 100.00  & 13.3  & 2.77 & \cite{Chen:2010qx,Fattoyev:2013yaa,Centelles:2010qh,Piekarewicz:2007dx,Ducoin:2010as,Fortin:2016hny} \cr
NL3$\omega\rho$ & 31.70  & 55.50  & -7.60  & 13.0  & 2.75 & \cite{Fortin:2016hny} \cr
Rs & 30.59  & 80.06  & -9.10  & 10.8  & 2.12 & \cite{Steiner:2004fi,Fortin:2016hny} \cr
SFHo & 31.60  & 47.10  &  & 10.5  & 2.05  & \cite{Oertel:2016bki} \cr
SFHx & 28.70  & 23.20  &  & 10.9  & 2.14  & \cite{Oertel:2016bki} \cr
SGI & 28.30  & 63.90  & -51.99  & 12.0  & 2.25  & \cite{Steiner:2004fi,Lim:2013tqa,Dutra:2012mb,Fortin:2016hny} \cr
SK255 & 37.40  & 95.00  & -58.30  & 11.0  & 2.15 & \cite{Fortin:2016hny} \cr
SK272 & 37.40  & 91.70  & -67.80  & 11.2  & 2.24 & \cite{Fortin:2016hny} \cr
SKa & 32.91  & 74.62  & -78.46  & 11.0  & 2.21  & \cite{RocaMaza:2011pm,Dutra:2012mb,Fortin:2016hny,Oertel:2016bki,Inakura:2015cla,Wang:2014rva} \cr
SKb & 33.88  & 47.60  & -78.50  & 10.6  & 2.2 & \cite{Fortin:2016hny,Dutra:2012mb} \cr
SkI2 & 33.40  & 104.30  & 70.60  & 11.3  & 2.17 & \cite{Ducoin:2010as,Fortin:2016hny,Dutra:2012mb,Inakura:2015cla} \cr
SkI3 & 34.80  & 100.50  & 72.90  & 11.4  & 2.25 & \cite{Ducoin:2010as,Fortin:2016hny,Dutra:2012mb,Inakura:2015cla} \cr
SkI4 & 29.50  & 60.40  & -40.60  & 10.6  & 2.18 & \cite{Ducoin:2010as,Fortin:2016hny,Dutra:2012mb,Inakura:2015cla,Lim:2013tqa} \cr 
SKI5 & 36.70  & 129.30  & 159.50  & 11.3  & 2.25 & \cite{Steiner:2004fi,Ducoin:2010as,Fortin:2016hny,Dutra:2012mb,Inakura:2015cla} \cr
SkI6 & 30.09  & 59.70  & -47.27  & 11.2  & 2.2 & \cite{Steiner:2004fi,Ducoin:2010as,Fortin:2016hny,Dutra:2012mb,Inakura:2015cla} \cr
SkMP & 29.89  & 70.31  & -49.82  & 10.8  & 2.11 & \cite{RocaMaza:2011pm,Ducoin:2010as,Steiner:2004f,Fortin:2016hny} \cr
Sly2 & 32.00  & 47.46  & -115.13  & 10.0  & 2.06 & \cite{Dutra:2012mb,Fortin:2016hny} \cr
Sly230a & 31.98  & 44.31  & -98.21  & 10.3  & 2.11 & \cite{Steiner:2004fi,Ducoin:2010as,Fortin:2016hny,Dutra:2012mb} \cr
SLy4 & 32.00  & 45.90  & -119.70  & 10.0  & 2.06 & \cite{RocaMaza:2011pm,Lim:2013tqa,Inakura:2015cla,Gonzalez-Boquera:2017rzy,Fortin:2016hny,Zhao:2016ujh} \cr
Sly9 & 31.98  & 54.86  & -81.42  & 10.6  & 2.16 & \cite{Dutra:2012mb,Fortin:2016hny} \cr
SV & 32.80  & 96.10  &  & 11.7  & 2.44  & \cite{Lim:2013tqa,Ducoin:2010as,Zhao:2016ujh,Dutra:2012mb} \cr
TFa & 35.05  & 82.50  &  & 12.2  & 2.10  & \cite{Fattoyev:2013yaa} \cr
TFb & 40.07  & 122.50  &  & 12.6  & 2.15  & \cite{Fattoyev:2013yaa} \cr
TFc & 43.67  & 135.20  &  & 12.9  & 2.21  & \cite{Fattoyev:2013yaa} \cr
TM1 & 36.90  & 110.80  &  & 12.5  & 2.18  & \cite{Ishizuka:2014jsa,Ducoin:2010as,Fortin:2016hny,Chen:2010qx} \cr
\hline
\end{tabular}
\end{table}

\twocolumngrid

\end{document}